\documentclass[preprint,prm,superscriptaddress]{revtex4-1} 
\usepackage{amsfonts,amssymb,amsmath,geometry, url,siunitx}
\usepackage{gensymb}
\usepackage{graphicx}
\usepackage{hyperref}
\usepackage{multirow}
\usepackage{setspace}
\hypersetup{
   colorlinks=true,    % false: boxed links; true: colored links
   linkcolor=blue,       % color of internal links
   citecolor=blue,       % color of links to bibliography
   filecolor=red,        % color of file links
   urlcolor=blue,        % color of external links
   linktoc=page          % only page is linked
}

\begin{document}

%\preprint{}

\title {Vacancy defect in bulk and at (10$\overline{1}$0) surface of GaN: A combined first-principles theoretical and experimental analysis}
\author{Sanjay  Nayak}
 %\email{sanjaynayak@jncasr.ac.in}
\affiliation{Chemistry and Physics of Materials Unit, Jawaharlal Nehru Centre for Advanced Scientific Research (JNCASR), Bangalore-560064}
  \author{Mit H. Naik}
  \author{Manish Jain}
  \affiliation{Centre for Condensed Matter Theory, Department of Physics, Indian Institute of
Science, Bangalore 560012}
 \author{U.V. Waghmare}%
% \email{Second.Author@institution.edu.}
\affiliation{Theoretical Sciences Unit \\ Jawaharlal Nehru Centre for Advanced Scientific Research (JNCASR), Bangalore-560064}
\author{S.M. Shivaprasad}
%\email{smsprasad@jncasr.ac.in}
\affiliation{Chemistry and Physics of Materials Unit, Jawaharlal Nehru Centre for Advanced Scientific Research (JNCASR), Bangalore-560064}   
%Jawaharlal Nehru Centre for Advanced Scientific Research (JNCASR), Bangalore-560064}

%

%\pacs{}% insert suggested PACS numbers in braces on next line

\date{\today}

\begin{abstract}
 We determine atomic and electronic structure, formation energy, stability and magnetic properties of native point defects, such as  Gallium (Ga) and  Nitrogen (N) vacancies in bulk and at non-polar (10$\overline{1}$0) surface of wurtzite  Gallium Nitride (\textit w-GaN) using, first-principles calculations based on Density Functional Theory (DFT). Under both Ga-rich and N-rich conditions, formation energy of   N-vacancies  is significantly lower than that of Ga-vacancies in bulk and at (10$\overline{1}$0) surface.  
 Experimental evidence of the presence of N-vacancies was noted from electron energy loss spectroscopy  measurements which further correlated with the high electrical conductivity observed in GaN nanowall network. We find that the Fermi level pins at 0.35 $\pm$0.02 eV below Ga derived surface state. Presence of atomic steps in the nanostructure due to formation of N-vacancies at the (10$\overline{1}$0) surface makes its  electronic structure metallic. Clustering of N-vacancies and Ga-Ga metallic bond formation near  these vacancies,  is seen to be another source of electrical conductivity of faceted GaN nanostructure that is observed experimentally.  
 
\end{abstract}

\pacs{ }% insert suggested PACS numbers in braces on next line
\maketitle 

\section{INTRODUCTION}
\setstretch{1.5}
Group III-nitride based semiconductors  are important  for use in opto-electronic devices\cite{Nakamura2011,Guha1998} such as light emitting diodes (LEDs) and lasers because  of  their direct and tunable band gap (0.7 eV to 6.0 eV). In addition,  GaN has  emerged as a strong candidate for dilute magnetic semiconductors \cite{Reed2001,Choi2005}, high power and high frequency devices \cite{Sheppard1999,Pearton2001}, and high electron mobility transistor (HEMT) applications \cite{Medjdoub2012}. As the group III-nitride semiconductors commonly crystallize in the polar wurtzite structure, they have a large internal piezoelectric field \cite{Langer1999}($\approx10^6  \ Vcm^{-1}$)  along (0001) direction, which supresses the radiative recombination. To avoid  such effect of piezoelectric fields on their electronic structure, GaN based hetero-structures are grown along non-polar directions like such as [10$\overline{1}$0] \cite{Wetzel2008} and [11$\overline{2}$0] \cite{Tanikawa2008}. During the growth process,  defects such as vacancies, dislocations nucleate naturally and are  commonly observed.
\par
 $\it{Ab-initio}$  calculations have been very effectively used   \cite{Stampfl1999,Gao2004,Gorczyca1999,Walle1994} in understanding the electronic properties of pristine \textit w-GaN as well as \textit{w}-GaN with defects.  Numerous attempts have been made to manipulate magnetism  in GaN by incorporating magnetic impurities\cite{reed2001room, lee2003magnetic,kronik2002electronic} for spintronic applications. Several authors proposed that the  cation vacancies in  \textit {w}-GaN lead to ferromagnetism \cite{Dev2008,Kuang2010} while,  anion vacancies in \textit{w}-GaN showed  paramagnetic behavior \cite{Larson2007}. This is liked with the fact that the cation vacancy  acts as  acceptor whereas, anion vacancy acts as  donor \cite{VanDeWalle2004}.
 While atomic and electronic structure of Ga vacancies in GaN have been studied extensively,  the structure and associated  properties of N vacancies remain elusive\cite{Li2010,Walle1994,Mattila1996,Gorczyca1999,Gao2004,Boguslawski1995,Xiao2008}.
 \par
 Overcoming the unintentional n-type doping of GaN has been  a great challenge for the semiconductor industry and the origin of such auto-doping is controversial.  Van de Walle and Neugebauer \cite{VanDeWalle2004} eliminated the N-vacancies as a possible cause of auto-doping, arguing that its formation energy is too high and suggested that oxygen and/or carbon impurity may explain the observed high conductivity in films grown with Metal Organic Chemical Vapor  Deposition (MOCVD), where organometallic precursors are used and could be one of the sources of the mentioned impurities. However, unintentional n-type doping and  high conductivity are also observed  in GaN grown with  Molecular Beam Epitaxy (MBE)\cite{Bhasker2012}. As MBE uses pure metal and ultrapure gas as the sources, and the films are grown in  Ultra High Vacuum (UHV) conditions, the cause of auto-doping in this case may be attributed to point defects instead of carbon and oxygen impurities.  Calculations carried out by Boguslawski \textit{et al.} \cite{Boguslawski1995} showed that N vacancy introduces a shallow donor state which might be relevant to auto-doping. 
\par
There is a lot of variation in the reported estimates of formation energy of N-vacancies in bulk \textit w-GaN under N rich conditions  based on $\it {ab-initio}$ calculations, which range from  1.1 to 5.08 eV\cite{Li2010,Walle1994,Mattila1996,Gorczyca1999,Gao2004,Boguslawski1995,Xiao2008}. Secondly, the  properties related to vacancies at the (10$\overline{1}$0)  surface\cite{Jin2009} have not been studied in depth. Defects can form more easily in lower dimensional structures, for example  at the surface of thin films and nanowires, due to the fact  that the surface itself is a planar defect having atoms with lower coordination numbers and  dangling bonds. 
\par
In the Refs. \onlinecite{Bhasker2012} and \onlinecite{Bhasker201572},  high electrical conductivity  was observed for the  GaN nanowall network (nanowall) grown on sapphire (0001) substrate and  the surface electronic structure  was proposed to be at the origin of its high conductivity. Microscopic imaging of the structure shows that the structure has different facets such as (10$\overline{1}$0), (10$\overline{1}$1) and (10$\overline{1}$2). To investigate the role of (10$\overline{1}$0) surface in yielding a high conductivity of the material, we simulated the atomic structure and electronic properties of  (10$\overline{1}$0) surface of \textit{w}-GaN  with point defects such as Ga and N vacancies by using \textit{ab-initio} Density Functional Theory (DFT). For completeness and a consistent comparison, we also obtained atomic structure and electronic properties of Ga and N vacancies in bulk \textit{w}-GaN.

\section{Computational Details}

Our calculations are based on density functional theory as implemented in ``Spanish Initiative of Electronic System with Thousands of Atoms'' (SIESTA) code \cite{Soler2001}. Local Density Approximated (LDA) functional parametrized by Ceperley and Alder \cite{Ceperley1980} was used for treating exchange and correlation energy. Norm-conserving pseudopotentials generated by the scheme of Troullier  and Martin \cite{Troullier1991} in the Kleinman - Bylander \cite{Kreibig1982}  form were used for ionic cores of Ga and N with valence electronic configuration $3d^{10} 4s^ 2 4p^1$ and $2s^2 2p^3$ respectively. 
Interaction between core and valence electrons  was included as nonlinear core correction (NLCC)\cite {louie1982nonlinear}. Valence electron wave functions were expanded by using a  combination of single zeta (SZ) and double zeta orbitals with polarization function (DZP). Hartree   potential and charge density were computed on a uniformly spaced grid with a resolution corresponding to kinetic energy cutoff  of 200 Ry. Brillouin zone of \textit w-GaN was sampled by a $\Gamma$ - centered $5\times5\times3$ mesh of k-points in the unit cell of reciprocal space\cite{Pack1977}. Positions of all the atoms were allowed to relax  to minimize energy using the conjugate gradient technique until forces on each atom were less than 0.02  eV/{\AA}. For Ga vacancy  in bulk GaN, we used a $4\times4\times2$ super-cell (128 atoms), which amount to a vacancy concentration of 1.56\%. To study the N-vacancies in bulk GaN, we used three different supercell such as $2\times2\times2$, $3\times3\times2$ and  $4\times4\times2$, which induces vacancy concentrations of 6.25\%, 2.76\% and 1.56\%, respectively.
In simulation of  stoichiometric (10$\overline{1}$0)   surface of GaN, we used a symmetric slab of 32 atoms . A vacuum space of $ \sim$ 12 {\AA}   was used to keep the interaction between the periodic images of the slab weak. We  used a $2\times2$ periodic-cell in the plane of the slab to simulate surface vacancies (Ga and N).  
Formation energy of defects  in bulk as well as at surfaces was calculated using Zhang-Northrup scheme \cite{Zhang1991},  given by
\begin{equation}\label{formationeqn}
 E_f = E_{tot} (V_{N \ or \ Ga}) - E_{tot} (pristine) + \Sigma n_i\mu_{i {(N \ or\ Ga)}}+ q (E_F+E_{VBM}-\Delta V_{0/b}) + E_{q}^{corr}
 \end{equation}
 where $ E_{tot} (V_{N\ or\ Ga})$ and $E_{tot} (pristine)$ are the total energies of the super-cell containing the N or Ga vacancy and the reference pristine  supercell respectively. $n_i$ and $\mu_i$ represent the number of vacancies and chemical potential of \ $ i^{th}$ species respectively. In this work, we have calculated the defect formation energy under both Ga rich and N rich conditions. Under N rich conditions $\mu_N$ is the energy of N- atom (obtained from the total energy $E_{tot}(N_2)$  of $N_2$ molecule, \textit {i.e.} $\mu _ N = \dfrac{1}{2} E_{tot}{ (N_2)} $), and the chemical potential of Ga is  calculated using the assumption of thermodynamic equilibrium, $\it {i.e.}$ $\mu_{Ga}+\mu_N = E_{GaN}[bulk]$; where $E_{GaN}[bulk]$ is the total energy of one formula unit of bulk \textit {w}-GaN. Similarly, for Ga rich conditions $\mu_{Ga}$ is the chemical potential of Ga atoms in bulk $\alpha$-Ga (\textit{i.e.} $\mu_{Ga}= \mu_{Ga}[bulk]$), and chemical potential of N atom ($\mu_N$) is calculated from thermal equilibrium condition. Under Ga rich conditions, the defect formation energy of  Ga and N vacancy are shifted by -$\Delta H_f$ and $\Delta H_f$ respectively,  as compared to N rich condition, where  $\Delta H_f$ is the formation enthalpy of \textit {w}-GaN and is estimated by the formula, $\Delta H_f = E_{GaN}[bulk]-\mu_{Ga}[bulk]-\dfrac{1}{2} E_{tot}{ (N_2)}$. The estimated value of $\Delta H_f [GaN] $ is -1.9 eV, which is in good agreement with earlier calculations based on LDA\cite{Gao2004, Gorczyca1999}.  $E_F$, $E_{VBM}$  stand for fermi energy and bulk valence band maximum of the pristine GaN respectively. $\Delta V_{0/b}$ is used for aligning the electrostatic potentials of  bulk and the neutral defective supercells and can be obtained by comparing of the electrostatic potentials in the bulk-like region far from the  neutral defect and in the pristine bulk calculation. We have used  the electrostatic correction term $E_{q}^{corr}$ for charged defect in the supercell by using the correction method proposed by Van De Walle \textit {et al.}\cite{freysoldt2009} as implemented in CoFFEE code\cite{naik2017coffee}. We have used the static dielectric constant ($\epsilon_s$) of GaN as 9.06 obtained from Ref\onlinecite{feneberg2014band} in estimation of $E_{q}^{corr}$ .  
\par
The thermodynamic transition level $(q_1/q_2)$ between two charge state of defects is estimated by the relation
\begin{equation}
(q_1/q_2)  =  \dfrac {E_{f}(^{q_1};E_F=0) - E_{f}(^{q_2};E_F=0)} {q_2-q_1}
\end{equation}
where ${E_{f}(^{q};E_F=0)}$ is the formation energy of the defect in the charge state q when the Fermi energy is at the VBM.
\par
Surface energy ($\sigma$) is calculated using the relation
\begin{equation}
\sigma = \dfrac{1}{2A} (E_{slab} - n \times {E_{GaN}[bulk]}) 
\end{equation}
where $ E_{slab}$ is the total energy of slab, ``n" is the number of formula units of  GaN in the slab and \textit A is the the area of surface unit cell in slab. In simulations of vacancies at the surface, corresponding atoms were  removed from the two surfaces of the slab.

\section{Results and Discussion}
\label{A}
\subsection{Atomic and Elecronic structure of  bulk \textit {w}-GaN}
\subsubsection {Pristine  bulk \textit {w}-GaN }
\begin{figure}
    \centering
    \includegraphics[scale=0.25]{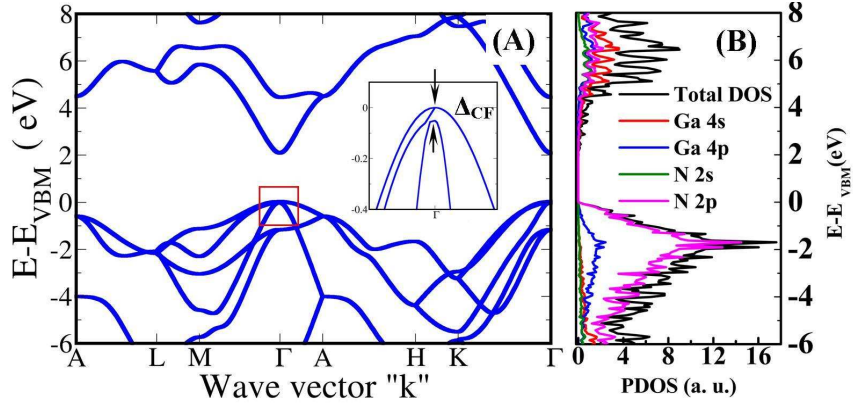}
     \caption{Electronic structure of bulk \textit{w}-GaN (A) and total DOS and PDOS are plotted in (B) obtained with SIESTA. (Color online)}
    \label{bulkband-dos}
   \end{figure}

 The equilibrium lattice parameters ($a$ and c), internal parameter (u) and the band gap (E$_g$) of bulk \textit w-GaN, estimated by different methods including experimental values are listed in Table \ref{tab1}. Our estimates of lattice parameters  agree well with earlier calculations and are within a  deviation of $\approx$ $0.4-0.5$\%  from the experimental values\cite{Paskova2006}. Our estimates of the band gap of bulk \textit w-GaN  is 2.06 eV at $\Gamma$ point (see Fig.\ref{bulkband-dos}(A)) is in good agreement with  earlier calculations\cite{Yan2014,Gonzalez-Hernandez2014} based on plane wave (PW) basis, but is underestimated \textit {with respect to} experimental  value of 3.4 eV at RT. Such   underestimation of band gap is typical of DFT-LDA  calculations \cite{Gonzalez-Hernandez2014,Wei1996,hybertsen1986electron,rubio1993quasiparticle} . 
Projected density of states (PDOS) (Fig.\ref{bulkband-dos}(B)) shows  that the valence band is primarily composed of  2p orbitals of N along with a small contribution from  4p orbitals of Ga while the conduction band is composed of 4s and 4p  orbitals of Ga, and a weak contribution from  2s and  2p orbitals of N. Our estimate of the crystal field splitting ($\Delta_{CF}$) (in the absence of spin-orbit coupling)  at $\Gamma$ point  is $\approx$ 50 meV, which is  smaller than the earlier theoretical estimates  of 72 meV by Suzuki \textit{et al.} \cite{Suzuki1995}, and a bit larger than 42 meV estimated by Wei \textit{et al.}\cite{Wei1996}, where both calculations were carried out  using LDA. The experimental estimates of $\Delta_{CF}$ reported in different works spanned from  10 to 25 meV\cite{Gil1995,Chuang1996,Layers1971,Reynolds1996}. Theoretical estimates of $\Delta_{CF}$ have been higher than its  experimental values.  The effective mass of electron near $\Gamma$ point (conduction band) is estimated  using  $\dfrac{1}{m^*}= \dfrac{1}{\hbar^2} (\dfrac{\partial^2E}{\partial k^2})$,  and our results are $0.16m_e$ (m$_\parallel^\ast , \Gamma - A $)  and 0.18$m_e$  (m$_\perp^\ast, M - \Gamma)$, in  reasonable  agreement with earlier  calculated estimates   ranging  from $0.18 m_e$ to $0.22 m_e$\cite{Suzuki1995,Gil1995,Chuang1996,Layers1971,Reynolds1996,Dreyer2013}, and  with the experimental values of $0.18 - 0.29 m_e$ \cite{Vurgaftman2003}.  Comparison of our results with the published results (see Table \ref{tab1}) benchmarks  the numerical parameters used in our work here, and also gives an idea of magnitude of errors.   
 
 \begin{table*}[!htb]
 \caption{\label{tab1}Optimized lattice parameters  ($a$ and c in {\AA}), internal parameter (u) and band gap ($E_g$ in eV)  of bulk \textit w-GaN } 
 \begin{ruledtabular}

  \centering
 \begin{tabular}{c c c c c c c rrrrrrr }
{ } & {SIESTA } &{SIESTA} &{VASP} &{VASP }  & {Expt.} &{Present }\\ 
  { } & {LDA } &{GGA} &{LDA} &{LDA}  & { } &{LDA }\\ 
 { } & {Ref. \onlinecite {Schmidt2002a}} &{Ref. \onlinecite{Carter2009}} &{Ref. \onlinecite{Yan2014}} &{Ref. \onlinecite{Gonzalez-Hernandez2014} }  & {Ref.\onlinecite{Paskova2006}}&{ }\\
\hline 
{$a$} & {3.23} &{3.28}&{3.155} & {3.160 }  & {3.189} & {3.17}\\
{c} & {5.19} &{5.31}&{5.145} & {5.150 }  & {5.186}& {5.16}\\
{u} & { -} &{0.378}&{0.3764} & {0.3765 }  & {0.377}& {0.377}\\
{$E_g$} & {2.37 } &{1.44}&{2.12} & {2.10 }  & {3.4}& {2.06}\\
%{b} & { - } &{- }&{- } & {-  } & {1.945} & {-}\\

\end{tabular}
\end{ruledtabular}
% \footnotetext[1]{Reference \onlinecite {Schmidt2002a}} 
% \footnotetext[2]{Reference \onlinecite{Carter2009}} 
% \footnotetext[3]{Reference \onlinecite{Yan2014}}
% \footnotetext[4]{Reference \onlinecite{Gonzalez-Hernandez2014}}
% \footnotetext[5]{Reference \onlinecite{Paskova2006}}
 \end{table*}                
 \subsubsection{Ga vacancies in bulk \textit {w}-GaN }
\begin{figure}[!htb]
    \centering
    \includegraphics[scale=0.5]{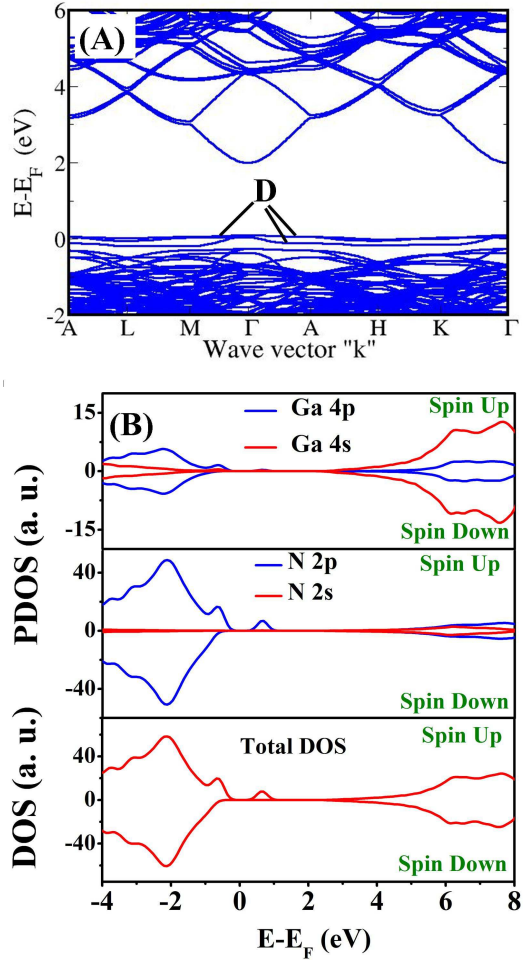}
    \caption{Electronic structure of bulk \textit{w}-GaN  with  Ga vacancy concentration of 1.56\% (A).Total DOS and PDOS are plotted in (B). (Color Online)}
    \label{ga-vac-bulk}
\end{figure}

Due to  a neutral cation (Ga) vacancies with   concentration of  1.56\%,  four neighbouring N atoms  move away from their positions causing a contraction of their bonds with other Ga  neighbors by $ \approx$ 1.9-2.3 \%. This  agrees  qualitatively but is  smaller than  earlier estimates of the change in bond-lengths of $ \approx$ 3.5-3.7\% reported by Neugebauer and Van de Walle, using a PW pseudopotential based calculation  with LDA approximation \cite{Walle1994} and by Carter and Stampfl \cite{Carter2009} using GGA approximation with SIESTA code (~2.9-3.7 \%). We find that Ga vacancies in charged state of -1, -2, and -3, the contraction in bond length varies from 2.5-2.6\%, 3.3-3.4 \% to 4.1-4.3\%, respectively.
Our estimate of the formation energy of neutral Ga vacancy  obtained under N (Ga) rich conditions is 6.90 (8.80) eV, which is  in good agreement with earlier calculations (see Table \ref{tab3}). The formation energy of  a Ga vacancy obtained under N (Ga) rich conditions at p-type growth condition  with -1$|e|$, -2$|e|$ and -3$|e|$ charge state is 7.17(9.07), 8.29 (10.19) and 10.18 (12.08) eV respectively. From the defect formation energy versus Fermi enegy plot (see Fig.\ref{transition-bulk} (a)), we find the thermodynamic transition levels (0/-), (-/2-) and (2-/3-) of $V_{Ga}$ are present at 0.27,  1.11 and 1.89 eV, above the VBM. Our estimate to the correction term ( $E_{q}^{corr}+q\Delta V_{0/b}$) is   0.12, 0.65, 1.56 eV for the defects with -1, -2 and -3$|e|$ charge state respectively. We find that  neutral Ga vacancy is more stable under p-type growth conditions while -3$|e|$ charged state is more stable under n-type growth conditions. The thermodynamic transition levels determined here  agree qualitatively with the calculation where finite size corrections are adopted\cite{lyons2017computationally}.  In the electronic structure of GaN with Ga vacancy concentration of 1.56\%  (see Fig.\ref{ga-vac-bulk}(A)), we identify the  defects  bands (denoted as D) by visualizing the spatial distribution of  wave functions at high symmetry k points such as $\Gamma$, A and M. Calculated electronic structure supports that the neutral Ga vacancies in bulk GaN are triple acceptors and the associated states are spin polarized. The three states are located 0.59, 0.63 and 0.63 eV above the valence band maximum (VBM) at $\Gamma$-point.               
As is evident in the spin dependent  DOS and PDOS of GaN with   vacancy concentration of 1.56\% (see Fig.\ref{ga-vac-bulk} (B)), acceptor states  have the 2p orbital (N) character,  and arise  from those N atoms which co-ordinate with the Ga vacancy,  displaying a local magnetic moment of $\approx$ 3.0 $\mu_B$, constistent with earlier report\cite{Dev2008} . The estimated spin polarization energy (E$_{spin\:  un-polarized}- E_{spin\: polarized}$) is  $\approx$ 0.2 eV, suggesting that the magnetic  state may be realized well above the room temperature. However, in the work of Dev \textit{et al.} \cite{Dev2008} the coupling between spins of the neutral Ga vacancies in bulk GaN is found to be Anti-Ferro Magnetic (AFM) in nature.  As the neutral Ga vacancies in bulk \textit{w}-GaN act as p-type dopants having weakly dispersed bands with a width ($\approx 0.4-0.7$ eV) and thus the observed higher conductivity of  GaN is unlikely to arise from  neutral  Ga vacancies. 
 
 \subsubsection{N vacancies in bulk \textit {w}-GaN }
To simulate  N vacancies at varying concentrations in bulk \textit {w}-GaN we used three  super-cells, $\it{ i.e.}$ $2\times2\times2$ (32 atoms), $3\times3\times2$ (72 atoms) and $4\times4\times2$ (128 atoms), and introduced a vacancy at N site amounting to a  vacancy concentrations of   6.25\%, 2.76\% and 1.56\% respectively. As stated earlier,  reports of defect structure and  formation energy of N vacancies in N-rich conditions are scattered over a wide range of values  \cite{Li2010,Walle1994,Mattila1996,Gorczyca1999,Gao2004,Boguslawski1995,Xiao2008},
 partly due to  two different relaxation processes  proposed by different groups. 
 Van de Walle \textit{et al.}\cite{VanDeWalle1994} reported, using plane wave DFT-LDA (32 atom supercell \textit{i.e.} vacancy concentration of 6.25\%), that the neighbouring Ga atoms move away from the N-vacancy site and found that the subsequent contraction of  Ga-N bonds near the vacancy is $\approx$ 4\%. This  is consistent with the calculations of  Gulan \textit{et al.} \cite{Gulans2005}, where the contraction of Ga-N bond length is by $\approx$ 3.7-3.9 \% (96 atom supercell) based on a GGA approximation of the DFT  and atomic orbital basis for the expansion of Kohn-Sham states. Further, Carter \textit{et al.} \cite{Carter2009}   reported a similar trend by using the GGA calculations with  SIESTA code, but their estimates of  bond length contraction is much weaker ($\approx$ 0.2-0.3 \% (96 atom supercell)) compared to the results discussed earlier. They concluded that such low values of outward relaxation may arise from  the inclusion of Ga 3d electrons in the valence.  Contrast to these results,  Gorczyca \textit{et al.} \cite{Gorczyca1999}  reported  inward relaxation of Ga atoms towards the N vacancy in 32 atoms supercell  of bulk \textit cubic phase of GaN (c-GaN) using the LMTO method (LDA)   and an elongation of Ga-N bond by $\approx$ 2\%. Another report by Chao \textit{et al.} \cite{Chao2007} based on GGA calculation with a plane wave basis code  also showed an inward relaxation of Ga atom causing elongation of Ga-N bonds by $\approx$ 1.9-3.4 \% using a 16 atom super-cell.  
 \begin{figure}[!htb]
    \centering
    \includegraphics[scale=2.2]{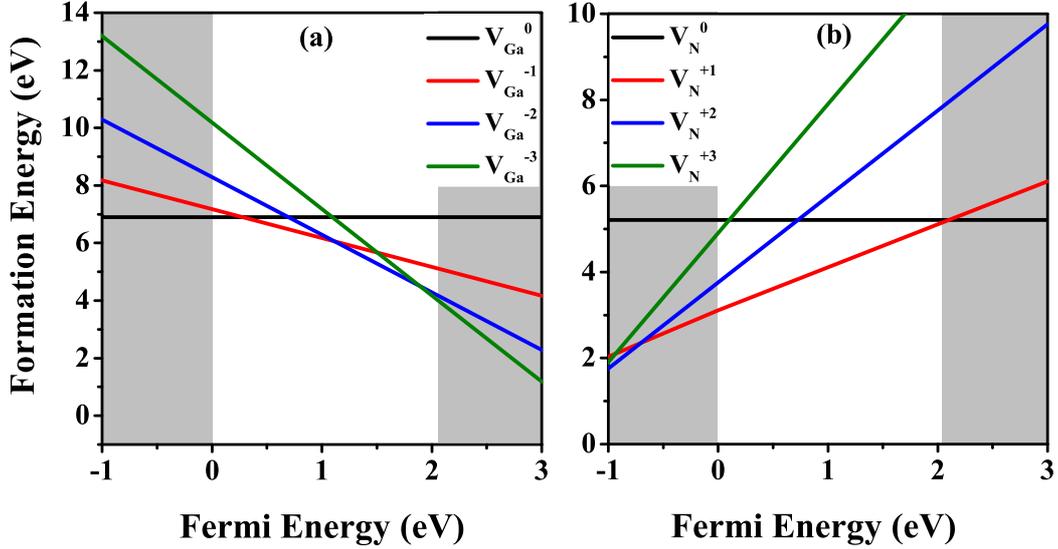}
    \caption{Defect Formation energy of Ga vacancy (a) and N vacancy (b) versus Fermi level in bulk. The white region indicate the LDA calculated band gap.}
    \label{transition-bulk}
\end{figure}
  To obtain a clear picture of ionic relaxtation due to  N-vacancies in \textit{w}-GaN, we have considered three different concentrations of N-vacancies : 6.25\%, 2.76\%, and 1.56\%. The structural changes due to ionic relaxation in these  configurations are  similar, and the Ga atoms that co-ordinate  with a N vacancy  move towards the  vacancy site causing elongation of their bonds with N neighbours.  The extent of elongation of bond-length varies with N-vacancy concentration. In the case of vacancy concentration of 6.25\%, 2.76\%, and 1.56\%, the Ga-N bond stretches by $\approx$ 1.23-2.5\%,  0.8-3.6 \% and 0.6-0.7\% respectively, similar to the theoretical predictions of  Gorczyca \textit{et al.} \cite{Gorczyca1999} and Chao \textit{et al.}\cite{Chao2007}. A recent work based on HSE methodology which is computationally much more expensive also predicts a similar trend\cite{yan2012}.   Such type of ionic relaxation has also been seen in the case of N vacancies in Indium Nitride (InN) \cite{Duan2008}.  For N vacancies in charged states  such as +1$|e|$, +2$|e|$ and +3$|e|$, the nearest neighbour Ga atoms displace away from the N-vacancy site causing reduction in Ga-N bond length to 0.10-0.35\% .  
\par
Formation energy of a N vacancy  estimated as a function of concentration is $\approx$  4.73, 4.75 and 5.21 and eV for $2\times2\times2$, $3\times3\times2$ and $4\times4\times2$ supercell respectively, under N-rich conditions. The  estimated formation energies of $V_N$ in p-type condition with charge state of +1$|e|$, +2$|e|$ and +3$|e|$ are 3.10, 3.75 and 4.90 eV respectively. The thermodymanic transition levels of  $V_N$ such as (+/0), (2+/3+), (3+/+) are present 2.10 eV above the VBM (\textit{i.e.} inside the conduction band), 0.7 and 0.9 eV below the VBM respectively. Our estimation of the correction term ( $E_{q}^{corr}+q\Delta V_{0/b}$) is   0.17, 0.74 and 1.70 eV for defects with +1$|e|$, +2$|e|$ and +3$|e|$ charged states respectively. We find that   $V_N$ in +1$|e|$ charge state is most stable throughout the band gap region within DFT-LDA (see Fig.\ref{transition-bulk} (b)).  
\par
\begin{figure*}[t]
    \centering
    \includegraphics[scale=0.5]{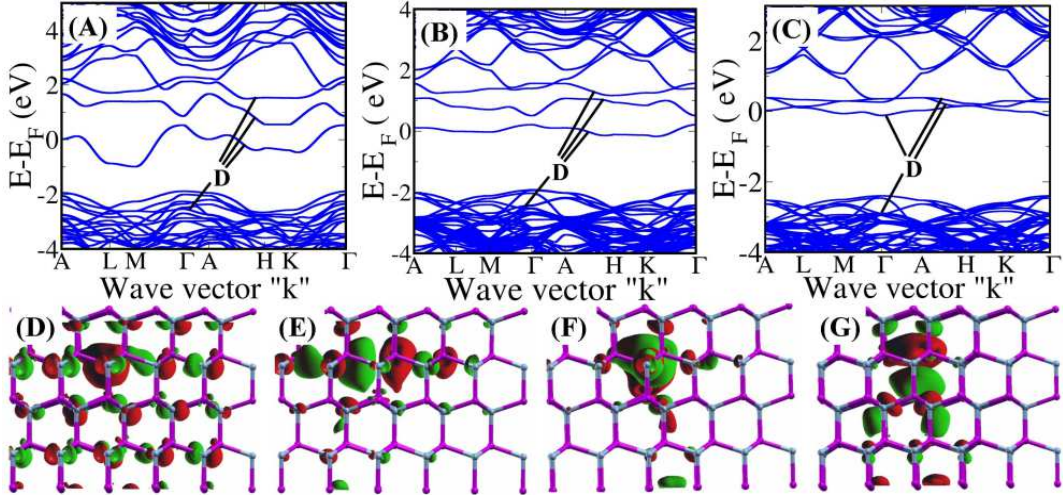}
    \caption{ (A), (B) and (C)  shows electronic structure  of N vacancies in bulk \textit{w}-GaN with vacancy concentrations of 6.25\%, 2.76\%, and 1.56\% respectively. All identified defect bands are designated as D. The wave function of defect bands that lie below the VBM and  the CBM are visualized and charge distribution of the bands are shown in (D), (E), (F) and (G) for the vacancy concentration of 1.56\%  respectively, with an isosurface value of $5.0\times10^{-2} e/\AA^3$.}
    \label{n-vac-bulk}
\end{figure*}

N vacancy in the bulk \textit w-GaN is a triple donor and introduces four defect states in the electronic structure.  Out of these defect states, as shown by Van de Walle \textit{et al.} \cite{Walle1994};  one fully occupied state lies below the conduction band and three other defect states lie inside the conduction band manifold. However, the work of Carter \textit{et al.} \cite{Carter2009}, where an outward ionic relaxation are noted, revealed the presence of  three singlet states near the CBM, two of which  are below the CBM by 0.1 and 0.9 eV respectively and third one is above the CBM by 0.3 eV. In addition, there is a state about 0.5 eV below the VBM. To find  the energies of defect states as a function of concentration  of N vacancy in GaN, we   examined their electronic structure and DOS  (see Fig.\ref{n-vac-bulk}).  At all  concentrations of N-vacancies  considered here one fully occupied state is present below the VBM and  the other three states are present near the conduction band minimum (CBM). We have visualized the bands at different high symmety k points and identified the defect bands  shown in Fig.\ref{n-vac-bulk}(D),(E),(F) and (G) for the N-vacancy concentration of 1.56\%. The energy level of fully occupied band lies below the bulk VBM  by  0.6 eV for defect concentration of 6.25 \% while for  2.76\% and 1.56\% this defect band lies 0.5 eV below the VBM. The other three states lie in the gap below the CBM. 
\par 
Further, to analyse the magnetic properties of N-vacancies in bulk \textit{w}-GaN,  we have obtained spin-polarized DOS within the LDA.  At all the defect concentrations studied here, we do not find any net magnetic moment. Our results are  consistent with the results of Xiong \textit {et al} \cite{xiong2009intrinsic}, while they differ from the other works \cite{Larson2007,Li2010} that reported  a net magnetic moment of 1.0$\mu_B$ per N-vacancy. To further-check our results, we  performed ``fixed-spin" calculation with a fixed  net magnetic moment of 1.0$\mu_B$, and found out  the N-vacancies with vanishing magnetic moment is  energetically  lower than with  a net magnetic moment of 1.0$\mu_B$ by 0.23 eV, at N-vacancy concentration of 6.25\%. Although, N-vacancy is   a triple  donor, the presence of a fully occupied state below the VBM makes it effectively a semiconductor with one electron per single N-vacancy for conduction. Higher defect formation energy and the presence of a fully occupied state just below the VBM indicate that the cause of auto-doping is not  the N vacancies in bulk \textit{w}-GaN. Secondly, very weak dispersion and band widths of the defect bands indicates that N vacancy states are localized and  may not contribute to  the observed high electrical conductivity in GaN nanowall.
\begin{table*}[!htbp]
 \caption{\label{tab:2} Calculated change in bond lengths  ($\Delta b$) w.r.t. bulk, vertical separation  between surface Ga and N atoms ($\Delta z$), buckling angle ($\omega_b$), surface energy ($\sigma$) and band gap ($E_g$) of ($10\overline{1}0$) surface of \textit{w}-GaN.}
 \begin{ruledtabular}
  \centering
 \begin{tabular}{c c c c c c  rrrrrrr }
{ } & {Present calculation } &{PWPP} &{VASP} &{VASP }  & {PWPP} \\
  { } & {LDA} &{LDA} &{LDA} &{GGA} &{LDA}  \\
 { } & { } &{Ref. \onlinecite {Northrup1996}} &{Ref. \onlinecite{Gonzalez-Hernandez2014}} &{Ref. \onlinecite{Gonzalez-Hernandez2014}}  & {Ref. \onlinecite{Filippetti1999}}\\
 \hline \
 {$\Delta b$ (in \%)} & {6} &{6}&{7.23} & {7.51 } & {6}  \\
{$\Delta z$(in {\AA})} & {0.44} &{0.22}&{-} & {- } & {0.36} \\
{$\omega_b$(in $ ^\circ$)} & { 14.03} &{7}&{7.5} & {8.183 } & {11.5} \\

{$\sigma$ (in meV/{\AA}{$ ^2$)}} & {168} &{118}&{123 } & {97.70} & {-}  \\
{$E_g$(in eV)} & {1.86} &{-}&{1.815 } & {1.534} & {-}  \\

\end{tabular}
\end{ruledtabular}
 \end{table*}                                                
     
  \label{B}
 \subsection{Atomic and electronic structure of   ($10\overline{1}0$) surface  of \textit {w}-GaN}
\subsubsection{Pristine ($10\overline{1}0$) surface }
\begin{figure*}[!htb]
    \centering
    \includegraphics[scale=0.8]{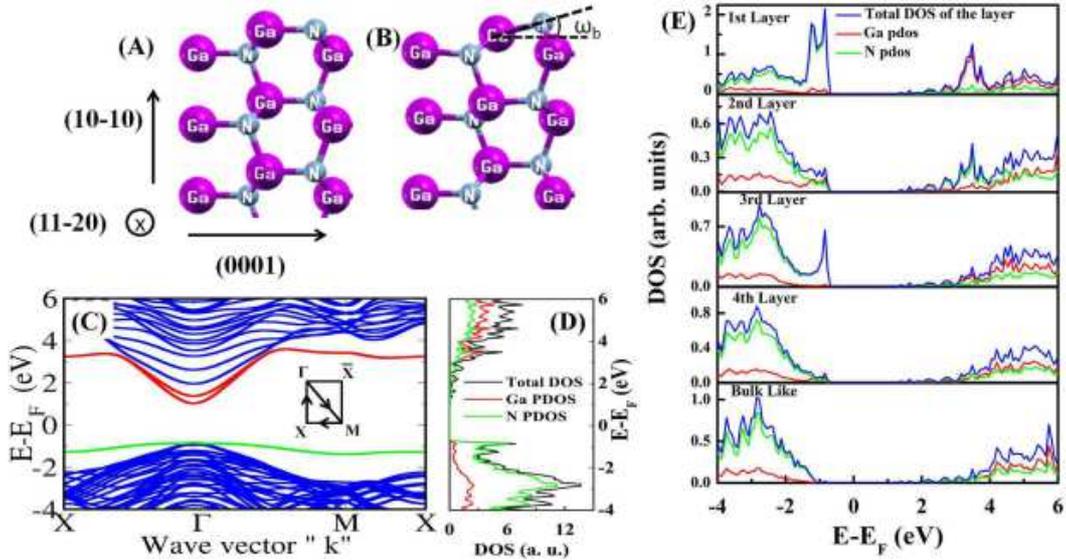}
    \caption{ (A) and (B) shows ideal and relaxed slab for ($10\overline{1}0$) surface respectively. (C) shows surface band structure of $1\times1$ slab of (10$\overline{1}$0) surface. The band colored in red denotes the surface Ga state and  in green represent surface N state. DOS along with PDOS of N (green) and Ga (red) atoms are shown in (D). Layer resolved total DOS (in black lines) of the slab with the projected DOS of N (in green) and Ga (red) atoms are shown in (E).} 
    \label{pristine-m}
\end{figure*}

(10$\overline{1}0)$ - surface of \textit {w}-GaN has  two configurations of surface termination, of which the one with a single dangling bond per atom  is energetically more stable than the other one with two  dangling bonds  at the surface \cite{Jindal2010}. This prompted the use of former in the present calculations, where a slab geometry is used to model the pristine non-polar (10$\overline{1}$0)  surface of GaN. In the relaxed  ($10\overline{1}0$) surface of \textit {w}-GaN  (shown in Fig.\ref{pristine-m}(B)),   Ga atoms  at the surface of the slab move inwards into the bulk, whereas  N atoms  move outward into the vacuum, causing a vertical separation of $\approx$ 0.4 {\AA} along $<10\overline{1}0>$ between Ga and N atoms and   buckling of surface Ga-N bond by $14.2^\circ$, which is slightly over-estimated than 7 - 11$^\circ$ obtained in calculations based on PW basis\cite{Gonzalez-Hernandez2014,Filippetti1999}. Upon structural relaxation at the surface, bond length of Ga-N  at ($10\overline{1}0$) surface  reduces to 1.83 {\AA} \textit{i.e.} contracted by $\approx$ 6\% \textit{w.r.t}  that in bulk. This  agrees well with earlier works \cite{Gonzalez-Hernandez2014,Northrup1996,Landmann2015,Northrup1996,Landmann2015,Filippetti1999}, where it was suggested that the structural relaxation of (10$\overline{1}$0) surface involves re-hybridization of surface Ga and N atoms resulting in $sp^2$ and $sp^3$ hybridization respectively\cite{Northrup1996}, which is also evident in our analysis. Our estimates of the surface energy ($\sigma$) of ($10\overline{1}0$) surface is 168 meV/{\AA}{$ ^2$} which is a bit overestimation in comparison to other published reports (see Table \ref{tab:2}).
\par
From the surface electronic structure (shown in Fig.\ref{pristine-m}(C) and (D)), a fundamental direct band gap of ($10\overline{1}0$) surface slab is  1.86 eV, which is 0.2 eV lower than the calculated bulk band gap. This is because of   the two surface states that arises within the band gap of bulk \textit w-GaN. The occupied surface N state ($S_N$), which mainly originates  from the 2p orbital of the N atoms, from 1$^{st}$ surface layer  and $2^{nd}$ sub-surface layer, with  a small contribution from atoms in the $1^{st}$ sub-surface layer. In the electronic structure it appears near to valence band and has a weaker dispersion with a band width  of 0.44 eV reflecting its confinement to surface. Unoccupied Ga surface state ($S_{Ga}$) originates mainly from 4s orbitals of the Ga atoms and from 1$^{st}$ layer  with a weaker contribution from 2nd layer of the slab. Unlike $S_N$,  $S_{Ga}$  has a higher dispersion with a band width of 2.2 eV and appears near to conduction band in the electronic structure. Layer resolved DOS (see Fig.\ref{pristine-m}(E)) clearly shows the origin of surface states, where a significant difference can be seen between DOS of the bulk-like layers and the first layer of the slab. From the spin polarized DOS  it is seen that the dangling bonds at ($10\overline{1}0$) surface do not necessarily show spontaneous spin polarization. The pristine ($10\overline{1}0$) surface is insulating in nature, and our estimate of the effective mass of electrons in Ga derived surface states is $\approx 0.2 m_e$ similar  to that of bulk. Thus high electrical conductivity may not arise from the conduction through these surface states either.\\

\textbf{Methodology used in the estimation of $E_q^{corr}$ for slab calculations} \\
 \begin{figure}[!htb]
    \centering
    \includegraphics[scale=0.35]{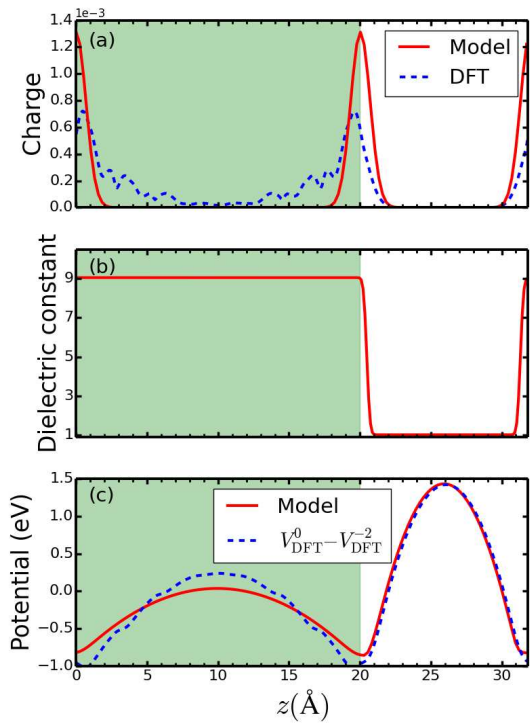}
    \caption{(a) shows charge density profile of defect state obtained from DFT calculation (dashed line) and model profile used for estimation of E$_{corr}$. (b) shows dielectric profile used for same. (c) shows the Hatree potential obtained from DFT calculation (dashed line) and with model charge distribution (solid line) }
    \label{model}
\end{figure}
\begin{figure}[!htb]
    \centering
    \includegraphics[scale=0.35]{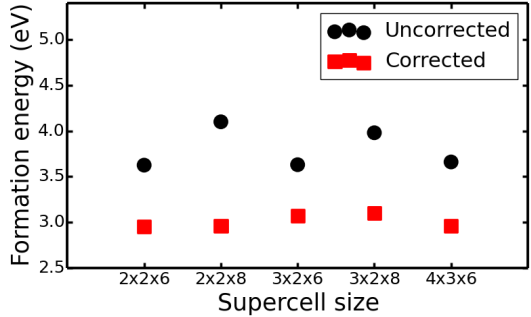}
    \caption{shows formation energy of gallium vacancy with -1 charge state with various supercell sizes. Note that for these calculation ionic relaxation were not performed.}
    \label{formation}
\end{figure}
Before we discuss  effects of Ga and N vacancies on structural and electronic properties of (10$\overline{1}$0) surface of GaN,  we provide a brief discussion on the methodology and reliability of the estimation of $E_q^{corr}$ term of Eqn. \ref{formationeqn}. The correction term is computed using the CoFFEE code by solving the Poisson equation for a model system. Figure \ref{model}(a) shows a plot of the planar averaged defect wavefunction. It can be seen that the defect states are localized at the surface, around the site of the defect. To obtain the electrostatic corrections, they are modelled using a Gaussian distribution as shown in Fig. \ref{model}(a). A model dielectric profile of the slab is constructed as shown in Fig. \ref{model}(b) to mimic the total charge density of the slab. The value of the dielectric constant inside the material is taken as 9.06 and outside the material as 1. Figure \ref{model}(c) shows the DFT difference potential, $V_{DFT}^{0} - V_{DFT}^{-2}$, and the model potential. They match well far from the defect, which indicates that the chosen model charge and dielectric profile are appropriate for this system. For the estimation of E$_{iso}$, we extrapolate the value of the model electrostatic energy (E$_{per}$) to infinitely large supercell. The electrostatic correction is then given by E$_{iso}$ - E$_{per}$. In Fig.\ref{formation}, we show the uncorrected and corrected formation energy of V$_{Ga}$ on the surface of the slab with -1 charge state, with varied supercell sizes and different vacuum dimensions. The thickness of the GaN slab is kept fixed in these calculations. The corrected formation energies are found to be the same (upto 0.1 eV) irrespective of the supercell size or the vacuum dimension.
\subsubsection{Ga-vacancies at ($10\overline{1}0$) surface } 
 In the simulations of Ga vacancies at the ($10\overline{1}0$) surface, we used a $2\times2$ in-plane super-cell (128 atoms) and introduced a vacancy  at  Ga site on both surfaces of the slab amounting to a surface Ga vacancy concentration of 25\% (surface Ga:N=0.75). Upon structural relaxation of the ($10\overline{1}0$) surface with Ga vacancies at the surface, we find that the neighboring N atoms move away from vacancy site causing  contraction of Ga-N bonds ($\approx$ 1.5 - 2\%)  in the neighborhood of vacancies compared to the ideal Ga-N bonds  at ($10\overline{1}0$) surface. A similar structural relaxation process (contraction in Ga-N bonds by $\approx$2.3-3.5\%\cite{Wang2010a} and $\approx$2.9-3.7 \%\cite{Carter2009} ) is reported in the case of Ga vacancies at ($10\overline{1}0$) side wall surface of GaN NWs. We find that the contraction in Ga-N bonds varies from 1.5-1.7\%, 2.0-4.81\% to 3.4-6.5\%  with the surface Ga vacancies in charged states of -1$|e|$, -2$|e|$ and -3$|e|$  respectively.  Our  estimate of the  formation energy of a Ga vacancy at ($10\overline{1}0$) surface under N-rich conditions is 3.97 eV, which is 2.93 eV less in comparison to the bulk and is a consequence of  lower coordination of atoms on the surface. Our estimate of the  formation energy of surface Ga vacancies in charged states of -1$|e|$, -2$|e|$ and -3$|e|$ are 4.35, 4.62 and 7.92 eV respectively. Formation energy versus Fermi energy plot (see Fig.\ref{transition-surface} (a))  shows  neutral Ga vacancies are more stable under p-type growth conditions while  Ga vacancies in -2$|e|$ charged states  is more stable under n-type conditions with a thermodynamic transition level (0/2-)  at 0.32 eV above N atoms derived surface state (or surface VBM).                        
  \begin{figure}[!htb]
    \centering
    \includegraphics[scale=0.75]{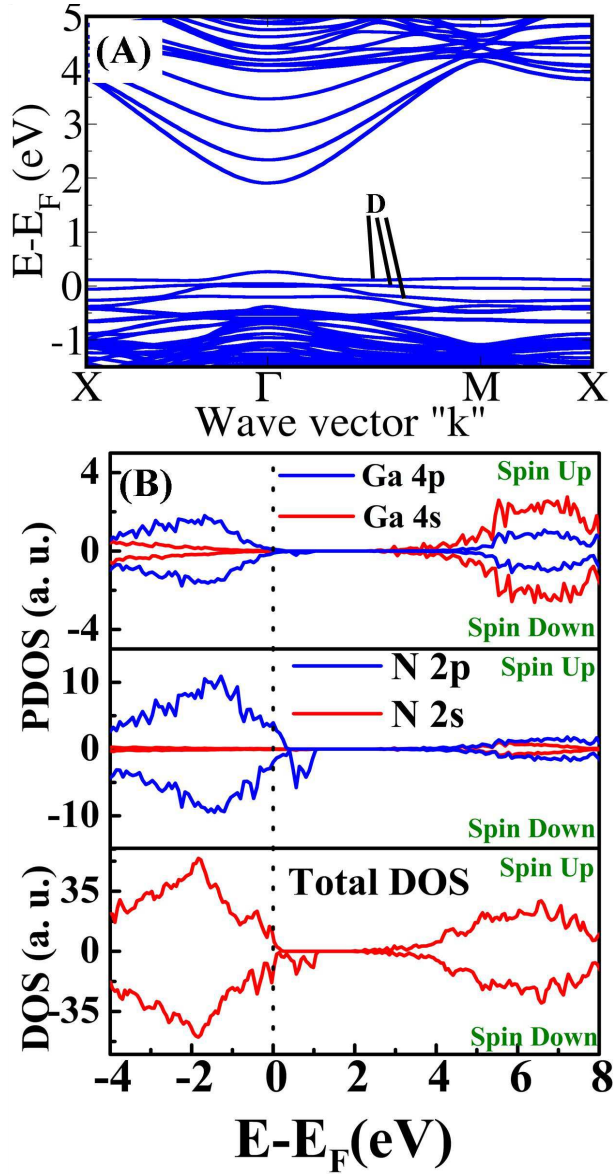}
    \caption{Spin un-polarized electronic structure (A) and spin polarized DOS and PDOS (B) of (10$\overline{1}$0) surface slab of w-GaN with  surface Ga vacancy concentration of 25\%. (Color Online)}
    \label{ga-vac-m-gan}
\end{figure}
\par
In the electronic structure  of ($10\overline{1}0$) surface with 25\% Ga vacancy concentration (see Fig.\ref{ga-vac-m-gan}), three vacancy-related states (designated as D) are evident near the valence band  just above the N atoms derived surface states. Thus, neutral Ga vacancies at ($10\overline{1}0$) surface act as a p-type dopant and haa a low band width suggesting that  the Ga-vacancies are not the source of the high electrical conductivity observed in GaN nanowall.  Furthermore, PDOS analysis (Fig.\ref{ga-vac-m-gan}(B)) shows  that these hole states originate from  2s/2p orbitals of the N atoms that co-ordinate with the Ga vacancy and are spin polarized. Due to symmetry breaking at the surface, the three N atoms contribute asymmetrically to the net magnetic moment with  individual magnetic moments of 1.462, 0.427 and 0.427 $\mu_B$  respectively. Estimated spin polarization energy of 1.1 eV is relatively higher than that for bulk GaN. Thus, magnetization  due to Ga vacancies at the surface is more stable than that in the bulk \textit{w}-GaN. From Fig.\ref{transition-surface} (a) it is clear that formation energy of the V$_{Ga}$ is too high to form in abundant  which can contribute to a significant number in free carrier concentration. Secondly, the obtained carrier type that is responsible for  electrical conductivity in the GaN nanowall is electron, thus we reiterate Ga vacancies are not responsible for the  high electrical conductivity observed in the GaN nanowall.

\subsubsection{N-vacancies at ($10\overline{1}0$) surface }         

\begin{figure}[!htb]
    \centering
    \includegraphics[scale=2.0]{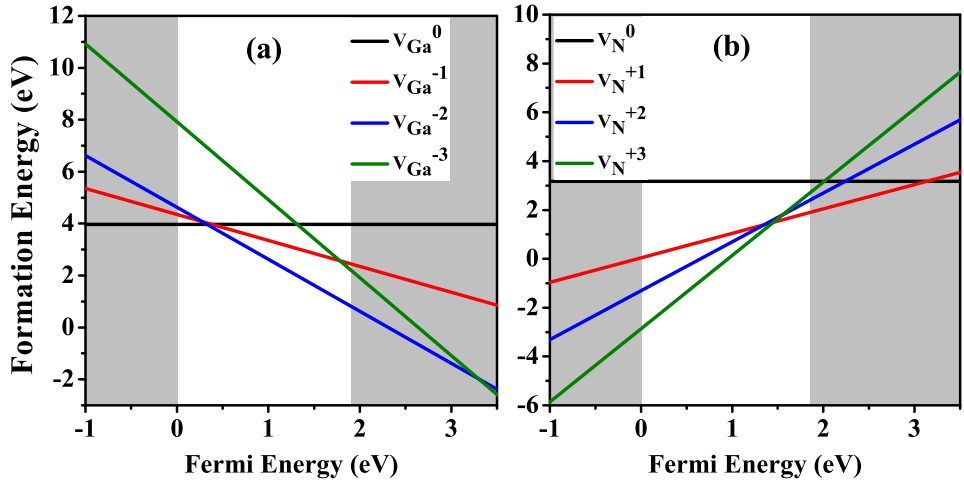}
    \caption{Defect Formation energy of Ga vacancy (a) and N vacancy (b) versus Fermi level on surface of the slab. The white region indicate the LDA calculated band gap.}
    \label{transition-surface}
\end{figure}

 To simulate N-vacancies at the ($10\overline{1}0$) surface, we introduced a vacancy at N site on each surface of the $2\times2$ in-plane super-cell of slab amounting to a surface N vacancy concentration of 25\% (surface Ga:N=1.33). We find upon structural relaxation of ($10\overline{1}0$) surface with N vacancies at surface, the Ga atoms sorrounding a N vacancy are found to move inwards the vacancy, resulting an elongation of Ga-N bonds by 2.6 - 6 \% compared to bond length at the pristine ($10\overline{1}0$) surface. Because of the N vacancies on ($10\overline{1}0$) surface, the basal plane Ga atom of vacancy site get displaced towards the vacuum, while remaing surface Ga atoms move inwards to bulk. Similar structural changes  have been reported by Carter \textit{et al.} \cite{Carter2009} in case of neutral N vacancies at the ($10\overline{1}0$) side wall surface of a GaN NWs. Further, in  +1$|e|$, +2$|e|$ and +3$|e|$ charged states of surface N vacancies, the stretching of  Ga-N bonds  varies from 0.64-1.56\%, 0.67-1.12\% to 0.48-0.51\% respectively compared to the ideal ($10\overline{1}0$) truncated surface. Our estimate of the formation energy of a N vacancy at ($10\overline{1}0$) surface of \textit w-GaN is  3.17 eV, which is 2.04 eV less than for bulk \textit w-GaN. Table ~\ref{tab3} shows the comparison of  formation energies of vacancies obtained in this work with those presented in the literature. The estimated  defect formation energy of a surface N vacancy in charges states  +1$|e|$, +2$|e|$ and +3$|e|$ are 0.04, -1.31 and -2.86 eV respectively. Formation energy versus Fermi energy plot (see Fig.\ref{transition-surface}) reveals that  N-vacancies in +3$|e|$ charged state is most stable under p-type growth conditions while +1$|e|$ charged state is more stable under n-type conditions with a thermodynamic transition level (3+/+) 1.45 eV above surface VBM.   It is clear that formation energy of the N-vacancy, both in bulk and at ($10\overline{1}0$) surface, is significantly less than that of  Ga vacancy  suggesting  the concentration of N-vacancies will dominate over that of Ga vacancies during the crystal growth. We can infer from Fig.\ref{transition-surface} (b) that N-vacancy can form spontaneously up to a Fermi level  $\approx$ 1.0 eV above surface VBM.  We further note that the Fermi level pins at 0.35$\pm0.02$ eV below the surface CBM.
 \par
\begin{figure}[!htb]
    \centering
    \includegraphics[scale=0.6]{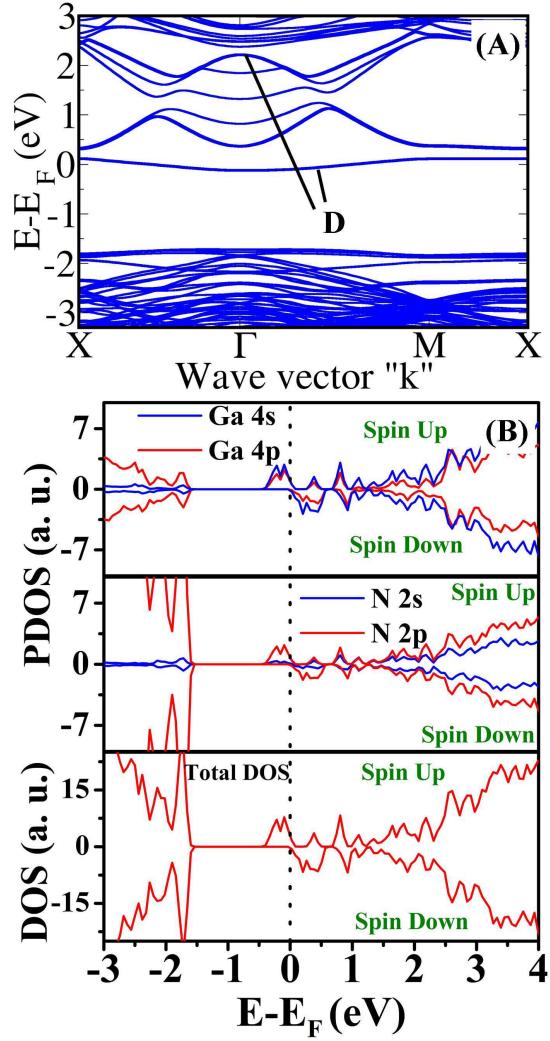}
    \caption{Spin un-polarized electronic structure (A) and spin polarized DOS and PDOS (B) of the ($10\overline{1}0$) surface of GaN with 25 \% of N vacancies. (Color Online)}
    \label{n-vac-m-gan}
    \end{figure}

Electronic structure of the ($10\overline{1}0$) surface with a N-vacancy  (see Fig.\ref{n-vac-m-gan}) at the surface reveals that they also act as n-type dopant and  donate only one electron per vacancy for conduction. Because of the N vacancy at ($10\overline{1}0$) surface, a half-occupied band appears in the fundamental band gap about 0.33 eV below the CBM at the $\Gamma$ point, while leaving the remaining defect bands to overlap with the conduction and valence bands. We find that the band width of the band near the Fermi level is $\approx$ 0.25 eV, and is spin polarized. In contrast to  N-vacancies in the bulk \textit{w}-GaN, asymmetry in spin polarized DOS (see Fig.\ref{n-vac-m-gan}(B)) can be seen near the Fermi level. From the spin polarized DOS, we find that the origin of this  state is in  4s/4p orbitals of Ga atoms that co-ordinate with the surface N vacancy and  2p orbitals of nearby N atoms.  A net magnetic moment of $\approx$1.0 $\mu_B$ per surface N vacancy is  estimated.  The Ga atom in the basal plane which is having 2 dangling bonds (db) due to the surface N vacancy contributes 0.415 $\mu_B$ each whereas the other two Ga atoms present in the 1$^{st}$ subsurface (with 1db)  contributes 0.1 $\mu_B$  to the total magnetic moment. Rest of the magnetic moment arises from the neighboring Ga and N atoms. A spin polarization energy of 0.11 eV is estimated, which is relatively lower than that of  surface Ga vacancy. Thus, relative stability of the magnetization  surface N-vacancies is weaker than than that of surface Ga vacancies. 
 
 \begin{figure*}[!htb]
    \centering
    \includegraphics[scale=0.65]{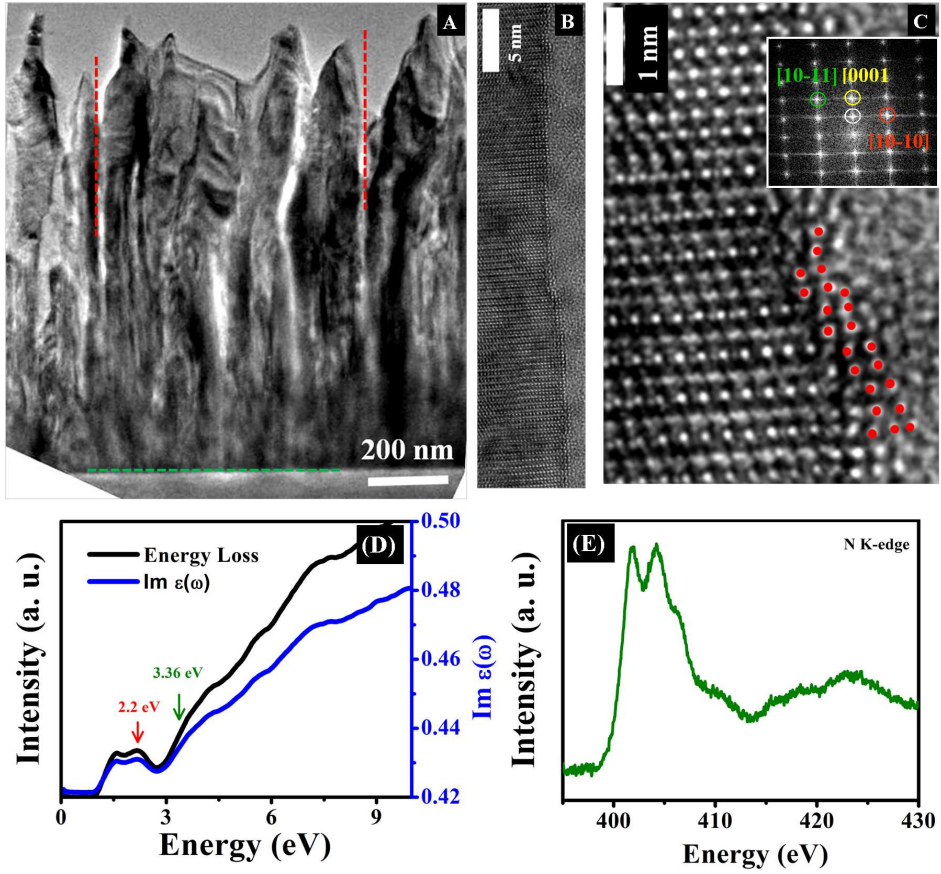}
    \caption{ Transmission Electron Microscopy (TEM) image of GaN nanowall network (nanowall) (A). (B) represent the High Resolution -TEM image of the side wall surface which is $(10\overline{1}0)$ surface. The formation of atomic steps can be seen in (C). (Color Online)}
    \label{TEM}
    \end{figure*} 
 
\begin{figure}[!htb]
   \centering
   \includegraphics[scale=0.6]{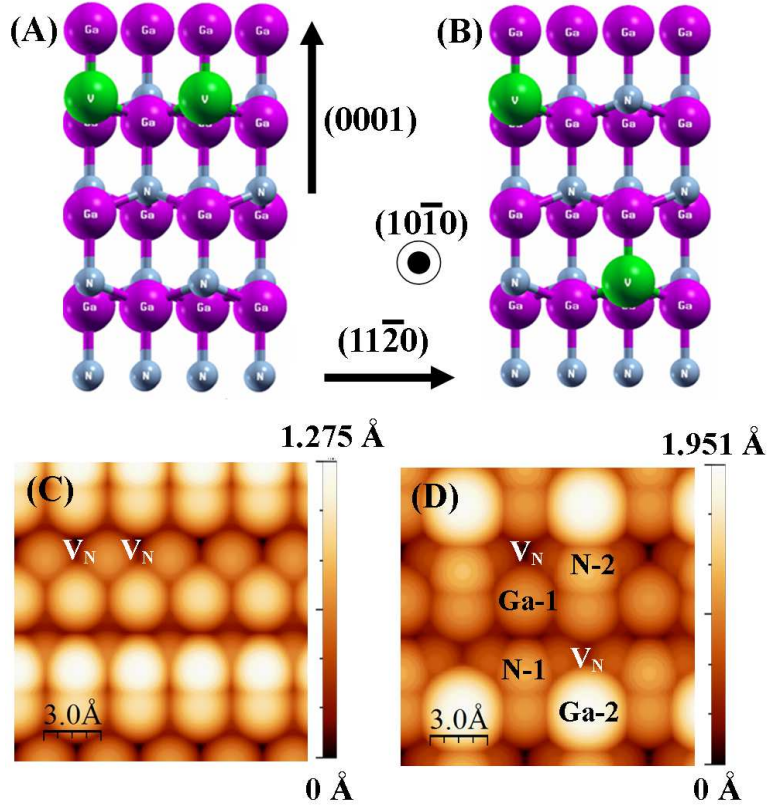}
   \caption{(A) and (B) shows the ideal atomic structure of vacancy configurations A and B respectively (Green spheres represent vacancy positions). (C) and (D) represent contour map representation of simulated STM image of surface configuration A and B respectively,  computed with WSxM software \cite{horcas2007wsxm} with constant current mode. The  scales represent the height of the surface atoms from the 1$^{st}$ subsurface layer.}
    \label{STM}
\end{figure}

Further, a closer look at the HR-TEM image of nanowall (see Fig.\ref{TEM} (A) and (B) reveals that the side wall surface makes an angle of $\approx$ 90\degree \space with (0001)-plane, with tapered surface at the top of the film due to formation of atomic steps (see red dots in Fig.\ref{TEM}(C)). FFT pattern of HR-TEM image (see Inset Fig \ref{TEM} (C)) confirms that the sidewalls consist of $(10\overline{1}0)$ surfaces of GaN. To simulate such atomic scale steps in the structure, we removed two N-atoms from each of  the surfaces of the slab, which gives rise to a surface N-vacancy concentration of 50\%. As  the slab is periodic and  extends in the XY plane, removal of two N-atoms from the surface is equivalent to removing a line of atoms creating a step. We considered two distinct configurations in the simulation of these  atomic step structure  (A and B) as shown in Fig. \ref{STM} (A) and (B). The separations between two N vacancies are 3.173 {\AA} and  6.06 {\AA} in A and B respectively. Total energies of  the  A and B structures reveal that  configuration A is more stable than configuration B by 0.25 eV per (2x2)  surface unit cell \textit{i.e.} proximal vacancies are more stable and promote the clustering of vacancies. 

\begin{table*}
\centering
 \caption{Vacancy formation energy for Ga and N vacancies in bulk \textit {w}-GaN  and at the ($10\overline{1}0$) surface of \textit{w}-GaN under both Ga and N rich conditions. All the values are given in electron volts (eV).}
 \label{tab3}
 \begin{ruledtabular}
\begin{tabular}{c|c|c|c|c|c|c|c|c rrrrrrrrr}
 {Configuration} & {charge  state} & \multicolumn{2}{|c|}{present calculation} & \multicolumn{2}{|c|}{Ref-1\cite{miceli2015energetics}} & {Ref-2\cite{laaksonen2008vacancies}} & {Ref-3\cite{kioseoglou2015energetic}} & {Ref-4\cite{limpijumnong2004diffusivity}} \\
\hline
\hline
                              &   & Ga-rich & N-rich  & Ga-rich  & N-rich & Ga-rich & Ga-rich & Ga-rich     \\
                               \hline
                 &&&&&&&&\\             
 V$_{Ga}^{Bulk}$ & 0  & 8.80 & 6.90& 7.02 & 6.58 &8.40  & 8.84  & 9.06   \\
                                & -1   & 9.07 & 7.17  & 8.90  & 8.46   & 8.83  & 9.23 & 9.31  \\
                                & -2    & 10.19  & 8.29  & 10.56  & 9.60 & 9.60&9.98 & 9.95        \\
                                & -3    & 12.08  & 10.18 & 14.14& 13.70 & 10.67 & 11.14  & 11.05  \\
                               
 V$_{N}^{Bulk}$   & 0 & 3.31& 5.21 & 2.59 & 3.03 & 3.16 & -&- \\
                                & +1 & 1.20 & 3.10& -0.58 & -0.14& 0.82 &- & 0.10 \\
                               & +2 & 1.85 & 3.75& -   & -  & 0.95 &  -  &-    \\
                                & +3    & 3.00 & 4.90 & -1.95  & -1.51  & 0.89   &- & -1.08   \\
 \hline
 &&&&&&&&\\
 V$_{Ga}^{Surf.}$ & 0  &5.87 &3.97  &-&-&-&-&- \\
                                & -1   & 6.25 &4.35&-&-&-&-&-   \\
                                & -2    & 7.34&4.62&-&-&-&-&-         \\
                               & -3    & 9.82&7.92&-&-&-&-&-  \\
                               
 V$_{N}^{Surf.}$ & 0  & 1.27 & 3.17  &-&-&-&-&- \\
                                & +1   & -1.86 &0.04&-&-&-&-&-   \\
                                & +2    & -3.21&-1.31&-&-&-&-&-   \\
                                & +3    & -4.76&-2.86&-&-&-&-&-  \\
 \end{tabular}
 \end{ruledtabular}
 \end{table*}

  \begin{figure}
   \centering
   \includegraphics[scale=0.15]{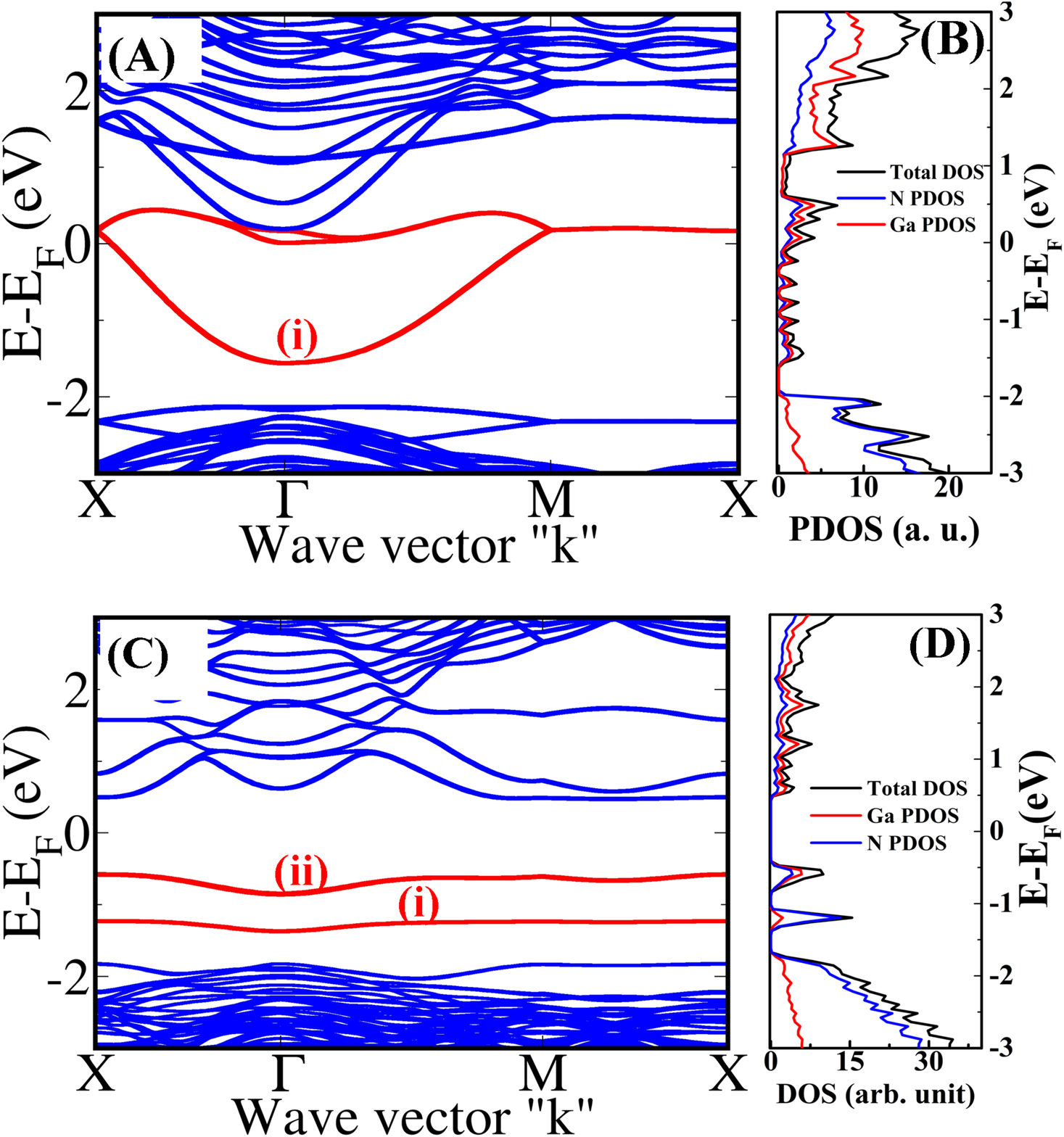}
   \caption{ (A) and (C) represent the surface electronic structure of configuration A and B respectively. Band colored in red are identified as defect  bands.  (B) and (D) represent DOS and PDOS of configuration A and B respectively.}
    \label{2n-vac-m-gan}
\end{figure}
\par
Ionic relaxation for configuration A shows that all the surface Ga atoms present  at the surface  moves along  (0001) direction and into the bulk region by 0.45 {\AA} from their ideal positions. The remaining N atoms  get displaced towards  vacuum. Such a displacement of Ga atoms  leads to the formation of Ga metallic clusters which include  Ga atoms of the first sub surface that are near the vacancy sites. The separation between surface Ga atoms and 1st subsurface Ga atoms near the N vacancies is  $\approx$ 2.5 {\AA}, which is very close to the  Ga-Ga bond length in bulk $\alpha$-Ga (2.534-2.818 {\AA}) \cite{Tonner2014}. Ionic relaxation of  configuration B   shows an unusual relaxation pattern and the simulated STM image obtained with constant current mode  is shown in Fig.\ref{STM} (D). From the STM image we can infer that the basal plane Ga atom (Ga-1)  moves towards the bulk \textit{i.e.} along [$10\overline{1}0$] and along [0001] direction, leading to the formation of Ga dimers in the subsurface. On the other hand, Ga atom (Ga-2)  (brightest one from the Fig.\ref{STM}(D)) move outwards to the vacuum. Out of two surface N atoms (N-1 and N-2 in Fig.\ref{STM}(D)), N-1 is displaced towards the bulk by 0.03 {\AA} while  N-2 moves towards the vacuum by 0.085 {\AA}, relative to the pristine ($10\overline{1}0$) surface. 
\par 
The surface  electronic  structure of A and B configurations is shown in Fig.\ref{2n-vac-m-gan} (A) and (C). In the electronic structure of configuration A, the Fermi level is in the conduction band indicating the metallic nature of the surface. The band designated as (i) in Fig.\ref{2n-vac-m-gan} (A) has a higher dispersion width of 1.71 eV with a minimum at X and a maximum at $\Gamma$ point in comparision to the defect bands in  N-vacancies in the bulk and at the ($10\overline{1}0$) surface having N vacancy concentration of 25\%.  PDOS analysis shows that the states (designated as (i) in Fig.\ref{2n-vac-m-gan} (A) and colored red),  arise from 4s/4p orbitals of Ga atoms with a weak additional contribution from  2p orbitals of nearby N atoms. Our estimate of electronic effective mass corresponding to band (i) is $0.5m_e$ (X-$\Gamma$). 
\par
Since the relaxed geometry of configuration B is different,  its electronic  structure exhibits a different trend in comparison with configuration A. The band denoted as (i) in Fig.\ref{2n-vac-m-gan} (C) is largely composed of N 2p orbitals of surface N atom N-1, while the same corresponding to N-2 appears near  VBM. The band marked with (ii) in Fig.\ref{2n-vac-m-gan} (C) is largely composed of 4s orbitals with a small contribution from 4p orbitals of surface Ga atom  (Ga-1) and 2p orbital of surface N atom (N-1). Band corresponding to remaining surface Ga atoms appears near the CBM. 
\par
From the above calculations, we note that formation energy of N-vacancies is comparatively less than Ga vacancies, thus the concentration of N vacancies is higher than Ga vacancies during crystal growth. However, the formation energy of N-vacancies in the bulk GaN is substantially high to form in abundance. Our calculations suggest that the N-vacancies at the surface can form spontaneously under p-type growth conditions indicating a large density of such defects at surface.  Further, the formation of steps by removing N atoms from  the surface of  ($10\overline{1}0$) slab shows  metallic character, thus we propose that the electrical conductivity observed in  Refs.\onlinecite{Bhasker2012} and \onlinecite{Bhasker201572} is probably  due to the  formation of atomic steps and a large number of nitrogen vacancies on the side wall surface of GaN nanowall,  giving rise to larger density of free electrons for conduction.  
\section{Experimental Validation}
 To validate the N-vacancies induced high electrical conductivity we probe the electronic structure of the material by EELS and they are shown in Fig.\ref{TEM}(D) and (E).  We have obtained both the Valence EELS and core-loss EELS of the $(10\overline{1}0)$ surface, where the beam direction is along  [$11\overline{2}0$] (Details of the EELS experiment can be found elsewhere\cite{dileep2016layer}). Fig.\ref{TEM}(D) shows the energy loss function and the imaginary part of the dielectric function (Im $\epsilon(\omega)$), obtained from Kramers-Kronig analysis. We have recorded a transition at 2.2 eV along with an early rise of dielectric profile (from $\approx$ 3.0 eV) in comparison to typical band gap of GaN (3.4 eV at RT). Observation of 2.2 eV peak in low loss EELS spectra indicate the presence of  deep and localized states in the gap which is quite consistent with the deep and localized character of N-vacancies in the bulk and slab calculations. The rise  of the dielectric profile before the energy of band to band transitions  elucidates that a large density of shallow donor states are also present. Such experimental results are quite consistent with the electronic structure  of N vacancy (both in bulk and slab) where unoccupied shallow states are being formed. The reduction in the electrical conductivity of GaN nanowall   after  chemical etching  also  suggests that the higher conductivity can be attributed to the localized nature of the defect state at the surface. X-ray Photoelectron Spectroscopy (XPS) study on a similar structure\cite{thakur2015electronic} reveals the formation of  Ga rich and n-type surface of the GaN nanowall due to N-vacancies, consistent with the Fermi level pinning of ($10\overline{1}0$) surface close to surface CBM, which supports our proposed mechanism.
\section{Summary}
 In summary, we have calculated  atomic and electronic structure, formation energy, stability and magnetic ground state  of native point defects  in bulk \textit w-GaN and at (10$\overline{1}$0) surface using first-principles DFT- based calculations. Ga vacancies, whose formation energy is significantly higher than N-vacancies under both Ga and N-rich condition, act as  p-type dopants and induce magnetism in GaN. N vacancy in GaN acts as n-type dopants (1 e/vacancy for conduction) and does not give rise to magnetic moment in the bulk, but a net magnetic moment $1.0\mu_B/$vacancy arises due to the N-vacancy at (10$\overline{1}$0) surface. We find that in bulk configuration, Ga vacancies in neutral state is more stable under p-type growth conditions while under n-type growth conditions it is more stable in -3$|e|$ charged state.  At (10$\overline{1}$0) surface, Ga vacancies in neutral state is more stable under p-type growth conditions while Ga vacancies in  -2$|e|$ charged state is more stable.  Further, in the bulk configuration, N vacancies in -1$|e|$ charged state is most stable  throughout the LDA estimated band gap region. At (10$\overline{1}$0) surface, N vacancies in +3$|e|$  charged state is more stable under p-type growth conditions  while  N vacancies in +1$|e|$ charged states is most stable under n-type growth conditions. Most importantly,  N-vacancies at the surface form spontaneously under p-type growth conditions giving rise to native n-type character of (10$\overline{1}$0) surface. Experimental evidence on the presence of N-vacancies was found by EELS measurements. Formation of an atomic step due  of N-vacancies on (10$\overline{1}$0) surface is found to give a metallic  electronic structure with the clustering of vacancies and Ga-Ga metallic bond formation near the vacancies, which we attribute to be responsible  for the observed  high electrical conductivity in  (10$\overline{1}$0) faceted GaN nanowall.   
\bibliographystyle{unsrt}
%\bibliography{diff} 

\end{document}